\documentclass[aps,prl,floats,twocolumn]{revtex4}

\usepackage[latin1]{inputenc}
\usepackage{graphicx}

\usepackage{amsmath,amsthm,amsfonts,eufrak,amssymb}
\input amssym.def
\newsymbol\square 1003
\newcommand{\al}{\alpha}

\begin{document}

\title{Hilbert's 17th problem and the quantumness of states}
\author{J. K. Korbicz$^1$, J.I. Cirac$^2$, Jan Wehr$^3$ and M. Lewenstein$^1$}
\affiliation{$^1$ Institut f\"ur Theoretische Physik, Universit\"at Hannover, D-30167
Hannover, Germany}
\affiliation{$^2$ Max-Planck Institut f\"ur Quantenoptik, Hans-Kopfermann Str. 1, D-85748, Garching, Germany}

\affiliation{$^3$ Department of Mathematics, University of Arizona, 617 N. Santa Rita Ave., Tucson, AZ 85721-0089, USA}

\begin{abstract}
A state of a quantum systems can be regarded as {\it classical} ({\it quantum}) with respect to measurements of a set of canonical observables iff there exists (does not exist) a well  defined, positive  phase space  distribution, the so called Galuber-Sudarshan $P$-representation. 
We derive a family of classicality criteria that require that averages of positive functions calculated using $P$-representation must be positive. For polynomial functions, these criteria are related to 17-th Hilbert's problem, and have  physical meaning of generalized squeezing conditions; alternatively, they may  be interpreted 
as {\it non-classicality witnesses}.  We show that every generic non-classical state 
can be detected by a polynomial that is a sum of squares of other polynomials (sos). We introduce a very  natural  hierarchy of states regarding their degree of quantumness, which we relate to the minimal 
degree of a  sos polynomial that detects them is introduced. Polynomial 
non-classicality witnesses can be directly measured.
\end{abstract}

\maketitle

In recent years there has been a lot of interest in classifying
states of quantum systems with respect to their quantum nature. In
particular, the problem of characterizing entangled states 
has attracted a lot of attention \cite{Bouw}
because of its vital importance for quantum information
processing. One of the aspect of this problem concerns
non-locality of quantum mechanics and violations of Bell-like
inequalities \cite{Bell} and existence of local hidden variable
models. The problem of existence of a {\it classical probabilistic
description} of quantum states of a {\it single system} has, however,
longer history and can traced back to the seminal papers of
Glauber and Sudarshan \cite{glauber}.

Let us consider a harmonic oscillator Hilbert's space, and fix the canonical creation and annihilation operators, $a$, $a^\dag$. In the Refs. \cite{glauber} it was shown that any state 
$\varrho$  can be uniquely put into a
form diagonal in coherent states $|\al\rangle$:
\begin{equation}
\varrho=\int d^2\alpha P(\alpha,\al^*) |\alpha\rangle \langle \alpha |\,,  \label{Prep}
\end{equation}
where $\al=x+\text{i}y$, $d^2\al=dxdy$. The integration with 
$P$ in (\ref{Prep}) is be understood in the
distributional sense \cite{Miller}. Hermiticity and normalization
of $\varrho$ imply that $P^{*}=P$ and $\int d^2\alpha
P(\alpha,\al^*)=1$, while positivity implies that $\int d^2 \alpha P(\alpha,\alpha^*) | \langle \alpha | \psi \rangle|^2\ge 0$ for every $\psi$. 

A state of a quantum systems is {\it classical} with respect to measurements of a given set of canonical observables iff the Glauber-Sudarshan $P$-representation is a well defined, positive  phase space distribution \cite{MandelWolf}.
Mathematically speaking,  in such a situation $P$ defines a
probabilistic measure $\mu$ on the phase-space through:
\begin{equation}\label{mju}
\mathbb{R}^2\!\!\supset\!\!\Omega\mapsto \mu(\Omega):=\int_\Omega P d^2\al \ge 0 \,.
\end{equation}
Statistical properties of state possessing the positive $P$-representation are as those of the classical statistical ensemble,
described by the measure $\mu$; that explains why such states are called {\it classical}.
However, the class of allowed $P$'s is larger than
that \cite{Miller}, and there exist {\it non-classical} states (such as squeezed, or Fock states) for which the 
latter integral does not always exist, or attains negative values. 

In this Letter we derive a family of classicality criteria that require that averages of positive functions calculated using $P$-representation must be positive. For polynomial functions, these criteria are related to 17-th Hilbert's problem, which in its simplest
form states that not every positive semidefinite polynomial  must be a  sum of squares (sos) of other polynomials \cite{Reznick}.  Our criteria have  physical meaning of generalized squeezing conditions, and may  be interpreted 
as {\it non-classicality witnesses} (in analogy to {\it entanglement witnesses}, \cite{primer}).  We show that every generic non-classical state 
can be detected by a sos polynomial. 

Let us begin by observing  that the set of probabilistic measures forms a convex
subset of the set of all $P$'s. The extreme points of this set are
point concentrated measures
$\{\delta^2(\al-\beta);\beta\in\mathbb{C}\}$, and the
decomposition into these points is unique \cite{simpleks}. Hence
the classical states form a generalized simplex $\Delta$. This is
the general feature of sets of probabilities for classical systems
\cite{Mielnik}. Therefore, geometrically the problem of distinguishing
between classical and nonclassical states amounts to the
operational description of the simplex of positive measures
$\Delta$ in the space of all $P$-distributions. We note that one
encounters closely related problem in the study of quantum
entanglement in multipartite systems (\cite{primer} and references
therein), with the difference that the convex subset of
classically correlated states is not a simplex. 

The solution of such stated problem was recently proposed by
Richter and Vogel \cite{vogel}. They studied the characteristic
function of $P$, i.e. its Fourier transform:
\begin{equation}
\hat P (\xi):=\int d^2\al P(\al,\al^*) \text{e}^{2\text{i}(\xi_i x-\xi_r y)}=\text{tr}\{\varrho :W(\xi):\}\, , \label{Phi}
\end{equation}
where $\xi=\xi_r+\text{i}\xi_i$ and $W(\xi)=\text{e}^{\xi
a^{\dagger}-\xi^* a}$ is the Weyl operator. In what follows the
hat,$\hat{\phantom{a}}$, will always denote Fourier transform. The
criterion detecting positive measures is then provided by
Bochner's theorem \cite{Simon2}:

{\it $\hat P$ is a Fourier transform of a probabilistic measure
iff $\hat P$ is of positive type, i. e. for each number $n$ and
all possible sets $\xi_1,\dots,\xi_n \in \mathbb{R}^2$ the
$n\times n$ matrix $\hat P_{ij}:=\hat P(\xi_i-\xi_j)$ is positive
semidefinite (psd).}

The further test of $\hat P_{ij}$ being psd for fixed
$\xi_1,\dots,\xi_n$ is carried out using determinant
criterion: a $n \times n$ matrix is psd iff determinants $D_k$,
$k=1\dots n$ of all of the principal submatrices are
non-negative. This finally leads to the hierarchy of
conditions: a state $\varrho$ is nonclassical iff there exist
$k>2$ (for $k=1$, $D_1=1$ due to normalization) and points
$\xi_1,\dots,\xi_k$ such that $D_k<0$.

Our solution of the classicality-quantumness problem 
follows also from the Bochner's theorem. Note
that the condition for $\hat P$ to be a function of positive type
can be equivalently rewritten as: $\hat P$ is a function of
positive type iff for all $\chi\in\mathcal{D}(\mathbb{R}^2)$,
$\int d^2\al d^2\beta \chi(\al)^*\hat P(\al-\beta) \chi(\beta)\ge
0$, where $\mathcal{D}(\mathbb{R}^2)$ is a space of smooth test
functions with compact support \cite{Simon2}. Using the
convolution theorem the last integral is equal to $\int d^2 \al
P(\al,\al^*)|\hat\chi(-\al)|^2$. From Fourier transform theory,
$\hat \chi$ can be analytically continued to a function from
$\mathcal{Z}(\mathbb{C}^2)$, the space of entire functions,
satisfying specific bonds \cite{Gelfand}, and every element of
$\mathcal{Z}(\mathbb{C}^2)$ is of that form \cite{Simon2,Gelfand}.
Hence we obtain the criterion for classicality \cite{uwaga}:

{\it $P$ defines a probabilistic measure iff
\begin{equation}\label{crit}
\forall {f\in\mathcal{Z}(\mathbb{C}^2)},\quad \int d^2 \al \, P \,
|f_{\mathbb{R}}|^2\ge 0\,,
\end{equation}
where $f_{\mathbb{R}}$ denotes the restriction of $f$ to $\mathbb{R}^2$.}

Our approach offers new insights into the problem, and connects it
to the methods used in the study of separability.  
From (\ref{crit}) we obtain that a state $\varrho$ is nonclassical iff
there exists a test function $f\in\mathcal{Z}(\mathbb{C}^2)$ such
that $\int d^2\alpha\, P \, |f_{\mathbb{R}}|^2 < 0$. Since
$f_{\mathbb{R}}$ is real-analytic this condition can be rewritten
as:
\begin{equation}
\text{tr}\{\varrho :|f_{\mathbb{R}}(a,a^{\dagger})|^2:\} < 0 \,, \label{witness}
\end{equation}
implying that the state is
nonclassical iff there exists an observable
$:|f_{\mathbb{R}}(a,a^{\dagger})|^2:$ detecting it. Geometrically, the condition $\int d^2\alpha \,P
\,|f_{\mathbb{R}}|^2 =0$ defines a hyperplane in the set of all
$P$ distributions and hence a state is nonclassical iff there is a
hyperplane separating it from the simplex $\Delta$. This is
essentially the same approach as the one used in the theory of
entanglement witnesses \cite{witnesses, optimization}. Therefore,
we propose to call the observable from expression (\ref{witness})
{\it nonclassicality witness}. 
The above approach can be generalized if we allow the test
functions $f$ to depend on the state $\varrho$ in question. Then,
the observable in equation (\ref{witness}) becomes a nonlinear
function of the state, and may be termed a {\it nonlinear
nonclassicality witness} (compare \cite{Otfried}).

In the current Letter we restrict the class of investigated states
$\varrho$ to those, for which $P$ can be evaluated on an arbitrary
real polynomial of $x,y$. The vector space of such polynomials
will be denoted by $\mathbb{R}[x,y]$. Since any polynomial can be
represented as a Fourier transform of appropriate sum of
derivatives of the Dirac's delta function, the sufficient
condition for that is that $\hat P$ is a smooth function. We
denote the space of such $P$'s by $\mathcal{P}$.

Note, that since the test function $f$ appearing in (\ref{crit})
is entire, it can be almost uniformly approximated by a sequence
of complex polynomials on $\mathbb{C}^2$. Hence
$|f_{\mathbb{R}}|^2=\text{lim}_{N\to\infty}(u_N^2+v_N^2)$ for some
$u_N,v_N\in\mathbb{R}[x,y]$. The almost uniform convergence on the
real plane allows us to interchange integration and taking the limit in
(\ref{crit}) \cite{Gelfand}. This leads to the main theorem of the present
Letter:

{\it A state $\varrho$ with $P\in\mathcal{P}$  is classical iff for every polynomial $v\in\mathbb{R}[x,y]$}
\begin{equation}\label{crit2}
\int\!\! d^2 \al \, P \, v^2 = \text{tr} \{\varrho :v(a,a^\dagger)^2:\}\ge 0
\end{equation}

In fact the criterion (\ref{crit2}) has already been used for a
long time for detecting some important classes of quantum states.
There are two examples of the application of (\ref{crit2}) known
to the authors. The first one is the test for higher order
quadrature squeezing \cite{Mandel}: $\varrho$ is squeezed to the
order $2k$ if there exists a phase $\phi\in [0,2\pi)$ such that
\begin{equation}
\sum_{l=0}^{k-1}\frac{1}{2^l}\frac{(2k)!}{l![2(k-l)]!}\langle : (\Delta E_{\phi})^{2(k-l)}:\rangle <0 \, , \label{sq2}
\end{equation}
with
$E_{\phi}:=a\text{e}^{-\text{i}\phi}+a^{\dagger}\text{e}^{\text{i}\phi}$,
$\Delta E_{\phi}:=E_{\phi}-\langle E_{\phi} \rangle $ and the
averages taken w.r.t. $\varrho$. Obviously, (\ref{sq2}) has the
form of a violation of (\ref{crit2}) (we can always substitute
$v^2$ there with finite sums of such terms) with the polynomial:
\begin{equation}
w_{2k}(x,y;\phi):=\sum_{l=0}^{k-1}\frac{1}{2^l}\frac{(2k)!}{l![2(k-l)]!}[d_{\phi}(x,y)]^{2(k-l)}\label{sq1}\, ,
\end{equation}
where
\begin{equation}
d_{\phi}(x,y):=2\Big[x-\big\langle\tfrac{a+a^\dagger}{2}\big\rangle\Big]\text{cos}\phi+2\Big[y-\big\langle\tfrac{a-a^\dagger}{2\text{i}}\big\rangle\Big]\text{sin}\phi\,.
\end{equation}
The witness $w_{2k}$ depends on the tested state $\varrho$ and hence is a nonlinear witness.

The second example is the test for sub-Poissonian statistics of $a^\dagger a$ (number squeezing): $\varrho$ is number squeezed if $\langle :(\Delta a^\dagger a)^2: \rangle <0\label{sub_P2}$. The corresponding nonlinear witness is
\begin{equation}
w_P(x,y):=(x^2+y^2-\langle a^\dagger a \rangle)^2\, . \label{sub-P}
\end{equation}

Note that both nonlinear witnesses (\ref{sq1}) and (\ref{sub-P})
are optimal in the sense that they are zero on the extreme points
of $\Delta$, as for any $|\alpha\rangle$ all the moments of
normally ordered deviations vanish.

From (\ref{crit2}), we observe that for any $v\in\mathbb{R}[x,y]$, $v^2$ 
is positive semidefinite ({\it psd}), and so is every polynomial which is sum of 
such terms (we call such polynomials {\it sos} polynomials). One may ask
if the converse is also true, i.e., if every psd polynomial is
sos? 

This problem has been known in mathematics under the name of
Hilbert's 17th problem. The  answer is, quite surprisingly,
negative: there are psd polynomials which are not sos
\cite{Reznick}. For the case of $3$ variables this happens for a degree
$m\ge 6$. However, the explicit examples of psd, but not sos
polynomials are rare and were found quite lately. It is also worth 
mentioning that the connection between Hilbert's 17th problem and
separability was established in \cite{choi}. 

In light of the
theorem (\ref{crit2}), out of all psd polynomials, sos polynomials
are sufficient to detect nonclassical states among the states with
$P\in\mathcal{P}$. To illustrate how the theorem (\ref{crit2}) works
let us consider a specific example of sixth order Motzkin polynomial which is psd, but non-sos:
\begin{equation}
M(x,y,z):=(x^2+y^2-3z^2)x^2y^2+z^6\label{Motzkin}\, .
\end{equation}
Using a method originating from the witness techniques in
entanglement study \cite{optimization}, we construct a state
$\varrho$, detected by the polynomial $M(x,y,\pm 1)$ \cite{dehom}.

Out of the four zeros $\{(\pm 1,\pm 1)\}$ of $M(x,y,\pm 1)$ we
construct coherent states: $\al_1:=1+\text{i}$, $\al_2:=-1+\text{i}$, $\al_3:=\al_2^*$, $\al_4:=\al_1^*$.
We pick the barycentric point, $\tilde \varrho$, of the face
$\mathcal{F}:=\text{conv}\{\delta (\alpha-\alpha_1),\dots,\delta
(\alpha-\alpha_4)\}$ ($\text{conv}$ stands for a convex hull) of
the simplex $\Delta$. Note, that the hyperplane defined by the
witness $:M(a,a^\dagger,\pm 1):$
\begin{equation}\label{h}
h_M:=\{P\in\mathcal{P};\int dx dy \,P(x,y)M(x,y,\pm 1) =0\}\, ,
\end{equation}
contains the face $\mathcal{F}\subset\Delta$ and hence the witness is optimal. Thus, to get the state detected
by (\ref{Motzkin}), we mix $\tilde \varrho$ with a projector onto
an arbitrary vector from it's range:
\begin{equation}
\varrho:=\frac{1-\epsilon}{4}\sum_{j=1}^4 |\alpha_j\rangle \langle \alpha_j |+\epsilon |\psi\rangle \langle \psi |\, , \label{ex}
\end{equation}
which for simplicity we choose to be:
\begin{equation}
\psi:=\frac{1}{N}(|\alpha_i\rangle+|\alpha_i^*\rangle)\, , \quad N^2=2[1+\text{e}^{-2}\text{cos}(2)]\,.
\end{equation}
Here $0\leq \epsilon \leq 1$ and $i\in\{1,2,3,4\}$ is fixed, but
the results presented below do not depend on it's particular
value. 
Calculating the average of the
polynomial (\ref{Motzkin}) using the expression (\ref{crit2}), we
obtain: $\langle:M(a,a^{\dagger},\pm 1):\rangle=(2/N^2)\,
\text{e}^{-2}\text{cos}(2)\, \epsilon$. Since $\text{cos}(2)<0$,
the state (\ref{ex}) is detected by $M$ for $\epsilon >0$.

As a side remark, we note that the state (\ref{ex}) is also
detected by another example of psd, but non-sos polynomial -
Choi-Lam polynomial $S(x,y,z):=x^4y^2+y^4z^2+z^4x^2-3x^2y^2z^2$,
as $\langle:S(a,a^{\dagger},\pm
1):\rangle=-(4/N^2)\text{e}^{-2}\text{sin}(2)\, \epsilon <0\,$.

Before we explicitly construct a sos polynomial detecting
(\ref{ex}), let us first examine the physically relevant witnesses
(\ref{sq1}) and (\ref{sub-P}). A simple calculation gives that
$\langle :(\Delta a^\dagger a )^2: \rangle\ge 0$ for any
$0<\epsilon \leq 1$.
Examination of the witnesses $w_{2k}$ is more difficult and we
have carried it out only numerically. We checked that up to the
14th order ($k=7$) all the inequalities (\ref{sq2}) are violated
for any $\phi\in [0,2\pi)$ and $0 < \epsilon \leq 1$ and hence
(\ref{ex}) is not squeezed up to the order of $14$. Apart from
that we used a modified version of $w_4$:
$d_{\phi_1}(x,y)^2d_{\phi_2}(x,y)^2+6d_{\phi_3}(x,y)^2$, depending
on three angles $\phi_1$, $\phi_2$ and $\phi_3$, also with no
success. 
The question if $\varrho$ has even higher order squeezing
is open.

To construct a sos polynomial detecting $\varrho$, note that  
$M(x,y,\pm1)$ has only four zeros, and
hence we can find a second order polynomial with the same zeros, 
which squared will give us the desired witness.
Equivalently, we look for such a
sos witness $W$ that its hyperplane $h_W$, defined as in (\ref{h}), contains
$\mathcal{F}$. We choose $W(x,y)=(Ax^2+By^2+Cxy+Dx+Ey+F)^2$. The
condition $W(\pm 1,\pm 1)=0$ leads to a system of four linear
equations for $A,\dots, F$. Its solution gives a family of
witnesses $W_{A,B}(x,y)=(Ax^2+By^2-A-B)^2$, where $A^2+B^2 \ne 0$. The average of $W_{A,B}$
in the state (\ref{ex}) is negative iff $\text{cos}(2)\big((A+B)^2-4A^2\big)+4\text{sin}(2)A(A+B)<0$. As this equation possesses non-zero solutions, for example $A=0,
B\ne 0$, the state $\varrho$ can be detected by a fourth order sos
polynomial.

This seems
to be a generic feature, at least for the psd polynomials of
degree $m=6$. In this case from \cite{Reznick} we know that if a
psd polynomial has exactly ten zeros in $\mathbb{PR}^3$, than it
cannot be sos. Fixing the variable $z$ generally reduced the
amount of zeros and hence permits to find a lower order sos
polynomial with the same zeros.

The methods described above, together with the criterion
(\ref{crit2}), can be used to classify the states according to the
degree of sos polynomial detecting them. Let us define a family of
convex subsets of $\mathcal{P}$:
\begin{equation}
S_m:=\bigcap_{w\in\tilde \Sigma_m}\{P\in\mathcal{P};\int d^2\al \, P \, w\ge 0\}\, ,
\end{equation}
where $\tilde \Sigma_m$ is the set of (inhomogeneous) sos polynomials of degree $m$. 
Theorem (\ref{crit2}) implies
that $\Delta =\bigcap_k S_{2k}$. It is also clear that $\tilde
\Sigma_2\subset\tilde \Sigma_4\subset\cdots$ and hence $S_2\supset
S_4\supset\cdots$. We prove a stronger result:

{\it For any even $m$ there exist nonclassical states detected by some
witness from $\tilde \Sigma_m$, and not by any witness from
$\tilde \Sigma_{(m-2)}$, that is $S_2\varsupsetneq S_4\varsupsetneq\cdots$.}

{\it Proof.} Let us choose a generic $w\in\tilde \Sigma_m$. It has
$(m+1)(m+2)/2$ terms, as it is a sum of polynomials of degree
$\le m$. From the variety $V(w):=\{(x,y);w(x,y)=0\}$ we pick $n$
points $(x_1,y_1),\dots,(x_n,y_n)$, $m(m+1)/2<n<(m+1)(m+2)/2$,
such that they do not lie on any variety of the lower order
$V(u)$, $u\in\tilde \Sigma_{(m-2)}$. We can find such points, as
otherwise there would exist $u\in\tilde \Sigma_{(m-2)}$, such that
$(x_1,y_1),\dots,(x_n,y_n)\in V(u)$. However, with chosen $n$ the
latter condition leads to an overcomplete system of linear
homogeneous equations for the coefficients of $u$, which
generically possesses no solution. On the other hand, same
condition for $V(w)$ yields under-determined system possessing
non-trivial solution. Having such points we construct coherent
states $|x_1+\text{i}y_1\rangle,\dots,|x_n+\text{i}y_n\rangle$ and
a face $\mathcal{F}_n\in\Delta$ spanned by them. For any $\tilde\varrho\in\mathcal{F}_n$ 
we have then that
$\text{tr}\{\tilde\varrho:w(a,a^\dagger):\}=0$, whereas
$\text{tr}\{\tilde\varrho:u(a,a^\dagger):\}>0$ for all $u\in\tilde
\Sigma_{(m-2)}$. Hence we can find such a convex combination
$\varrho$ of $\tilde\varrho$ and a projector onto some linear
combination of
$|x_1+\text{i}y_1\rangle,\dots,|x_n+\text{i}y_n\rangle$, such that
$\text{tr}\{\varrho :w(a,a^\dagger):\}<0$, while for all
$u\in\tilde \Sigma_{(m-2)}$,
$\text{tr}\{\varrho:u(a,a^\dagger):\}\ge 0$ (from the
continuity).$\,\square$

Summarizing, we have derived a family of classicality criteria of states of a quantum system, that require that $P$-representation averages of positive
functions are positive. For polynomial functions,  we have related these criteria to
17-th Hilbert's problem: we have proven the theorem that all "generic" non-classical states (for which the
$P$-representation averages of polynomials exist), can be detected
by sos polynomials of sufficiently high degree; in this sense non-sos polynomials (whose existence was proven by Hilbert) are not necessary for classicality detection. 
We have also introduced
the hierarchy of states implied by this theorem, and have introduced 
convex sets $S_2\varsupsetneq S_4 \varsupsetneq \ldots$ of  
states detected by squares 
of polynomials of the 1st, 2nd,... order, 
and corresponding in this sense to decreasing degree of quantumness. 

We stress that our results have important experimental consequences. 
Our polynomial {\it nonclassicality witnesses} can be easily measured, 
allowing thus for direct detection of quantumness and its 
degree for a given state. In this sense 
they are similar  to {\it entanglement witnesses}
that are nowadays commonly used for detection of entanglement \cite{demar}.
If one wants to check if a given state $\varrho$ 
is quantum, it is enough to measure
normally ordered averages of squares of real polynomials of position $q$, and momentum $p$, or 
quadrature operators. In order to check  the degree of quantumness (i.e. 
to check whether $\varrho \in S_{2k}$),
one should  determine normally ordered 
averages  of squares of real polynomials of the order $k$.
Note, that for a given $k$ this requires measurements of 
finite number of averages only. For instance, for $k=1$  (squeezing), 
one needs to measure $\langle q \rangle$, $\langle p \rangle$, $\langle :q^2:
 \rangle$,  $\langle :p^2: \rangle$, and   
  $\langle :qp+pq: \rangle$, and check if there exist $A,B,C$ such that
$\langle : (Aq+Bp+C)^2:\rangle<0$. For general  $k$, one needs 
respectively $k(2k+3)$ measurements. Our results for the first time fully 
categorize states with respect to their degree of quantumness,
and generalize the concepts of (higher order) squeezing or number squeezing 
as a signature of quantumness.

\acknowledgements

We acknowledge support from the Deutsche Forschungsgemeinschaft
(SFB 407, SPP 1078, GK 282, 436 POL) and the EU Programme QUPRODIS.


\begin{thebibliography}{99}
\bibitem{simpleks} Note that $\delta$ is a unique unit element w.r.t. convolution $P(\al,\al^*)=\int d^2\beta P(\beta,\beta^*)\delta^2(\al-\beta)$.
\bibitem{dehom} Any homogeneous polynomial
from $\mathbb{R}[x,y,z]$ can be dehomogenized by setting $z=1$, and vice versa: 
by introducing a third variable $z$, generic $v\in\mathbb{R}[x,y]$
can be homogenized. The
property of being psd or sos is conserved under de- and
homogenization.
\end{thebibliography}
\end{document}